\magnification \magstep1
\raggedbottom
\openup 1\jot
\voffset6truemm
\headline={\ifnum\pageno=1\hfill\else
\hfill {\it New Results in One-Loop Quantum Cosmology}
\hfill \fi}
\rightline {\ hep-th/9707138}
\vskip 1cm
\centerline {\bf NEW RESULTS IN ONE-LOOP QUANTUM COSMOLOGY}
\vskip 1cm
\leftline {Giampiero Esposito$^{1,2}$ and Alexander Yu.
Kamenshchik$^{3}$}
\vskip 0.3cm
\noindent
$^{1}$Istituto Nazionale di Fisica Nucleare, Sezione di
Napoli, Mostra d'Oltremare Padiglione 20, 80125 Napoli, Italy
\vskip 0.3cm
\noindent
$^{2}$Dipartimento di Scienze Fisiche, Mostra d'Oltremare 
Padiglione 19, 80125 Napoli, Italy
\vskip 0.3cm
\noindent
$^{3}$L. D. Landau Institute for Theoretical Physics of Russian
Academy of Sciences, Kosygina str. 2, Moscow 117334, Russia
\vskip 1cm
\noindent
{\bf Abstract.} A crucial problem in quantum cosmology is a 
careful analysis of the one-loop semiclassical approximation 
for the wave function of the universe, after an appropriate
choice of mixed boundary conditions. The results for Euclidean
quantum gravity in four dimensions are here presented, when
linear covariant gauges are implemented by means of the 
Faddeev-Popov formalism. On using $\zeta$-function regularization
and a mode-by-mode analysis, one finds a result for the
one-loop divergence which agrees with the Schwinger-DeWitt
method only after taking into account the non-trivial effect
of gauge and ghost modes. For the gravitational field, however,
the geometric form of heat-kernel asymptotics with boundary 
conditions involving tangential derivatives of metric perturbations
is still unknown. Moreover, boundary effects are found to be
responsible for the lack of one-loop finiteness of simple
supergravity, when only one bounding three-surface occurs. This
work raises deep interpretative issues about the admissible 
backgrounds and about quantization techniques in quantum
cosmology.
\vskip 100cm
A careful consideration of boundary conditions is known to
play a key role in several branches of classical and quantum
field theory: the problems of electrostatics, the theory of
vibrating membranes, the Casimir effect, the theory of van der
Waals forces, models of quark confinement, the path-integral
approach to quantum gravity and the quantum state of the
universe. Indeed, a rigorous definition of the Feynman sum
over all Riemannian four-geometries with their topologies does
not yet exist. However, the choice of boundary conditions for
metric perturbations and ghost modes may lead to a well defined
elliptic boundary-value problem, and this may be applied to
the one-loop semiclassical analysis of the quantum theory. 
Hence one may obtain a thorough understanding of the first set
of quantum corrections to the underlying classical theory,
which is a highly non-trivial achievement (despite the well
known lack of perturbative renormalizability of Einstein's
gravity). Such investigation leads, in turn, to a better
understanding of mixed boundary conditions and of the effective
action formalism in quantum field theory. Further motivations
result from the quantization of closed cosmological models,
and from the need to understand the relation between different
approaches to quantum field theories in the presence of
boundaries (e.g. reduction to physical degrees of freedom
before quantization, or the Faddeev-Popov gauge-averaging
method, or the extended-phase-space Hamiltonian path integral
of Batalin, Fradkin and Vilkovisky). At this stage, one can 
indeed anticipate a striking result: in a mode-by-mode 
evaluation of the one-loop wave function of the universe, 
including gauge-averaging and ghost modes, after performing
a 3+1 split and a Hodge decomposition of the components of
metric and ghost perturbations, there are no exact 
cancellations between contributions of gauge and ghost modes. 
This lack of cancellation turns out to be essential to obtain
agreement between different techniques [1].

We have first studied Luckock-Moss-Poletti boundary conditions
on metric perturbations $h_{\mu \nu}$ and ghost perturbations
$\varphi_{\mu}$ (hereafter, $\mu,\nu=0,1,2,3$, $K$ is the
extrinsic-curvature tensor, $g$ is the background metric, 
$n$ is the normal to the boundary, and
$P_{\mu \nu} \equiv g_{\mu \nu}-n_{\mu}n_{\nu}$):
$$
[h_{ij}]_{\partial M}=[h_{0i}]_{\partial M}=0 \; ,
\eqno (1)
$$
$$
\biggr[(2{\rm Tr}K+n^{\sigma}\nabla_{\sigma})
n^{\mu}n^{\nu}(h_{\mu \nu}-{1\over 2}g_{\mu \nu}
g^{\alpha \beta}h_{\alpha \beta})\biggr]_{\partial M}=0 \; ,
\eqno (2)
$$
$$
[\varphi_{0}]_{\partial M}=0 \; ,
\eqno (3)
$$
$$
\biggr[(-K_{\mu}^{\; \; \nu}+\delta_{\mu}^{\; \; \nu}
n^{\rho}\nabla_{\rho})P_{\nu}^{\; \; \sigma}\varphi_{\sigma}
\biggr]_{\partial M}=0 \; .
\eqno (4)
$$
These boundary conditions involve a mixture of Dirichlet and
Robin boundary conditions, but unfortunately are not completely
invariant under infinitesimal diffeomorphisms on metric
perturbations. For this purpose, one has instead to consider
the Barvinsky scheme:
$$
[h_{ij}]_{\partial M}=0 \; ,
\eqno (5)
$$
$$
[\Phi_{0}(h)]_{\partial M}=[\Phi_{i}(h)]_{\partial M}=0 \; ,
\eqno (6)
$$
$$
[\varphi_{0}]_{\partial M}=[\varphi_{i}]_{\partial M}=0 \; ,
\eqno (7)
$$
where $\Phi_{0}$ and $\Phi_{i}$ are the normal and tangential
components of the gauge-averaging functional. In linear covariant
gauges, e.g. de Donder, Eqs. (6) lead to normal and tangential
derivatives of $h_{00}$ and $h_{0i}$ [1]. 

Last, we have considered Robin-like boundary conditions 
on $h_{ij}$:
$$
\left[{\partial h_{ij}\over \partial \tau}+{u\over \tau}
h_{ij} \right]_{\partial M}=0 \; ,
\eqno (8)
$$
jointly with
$$
[h_{00}]_{\partial M}=[h_{0i}]_{\partial M}=0 \; ,
\eqno (9)
$$
$$
\left[{\partial \varphi_{0}\over \partial \tau}
+{(u+1)\over \tau}\varphi_{0} \right]_{\partial M}=0 \; ,
\eqno (10)
$$
$$
\left[{\partial \varphi_{i}\over \partial \tau}
+{u\over \tau}\varphi_{i} \right]_{\partial M}=0 \; .
\eqno (11)
$$
The resulting one-loop divergences on a portion of flat 
Euclidean four-space bounded by a three-sphere turn out to
be, in the de Donder gauge [1],
$$
\zeta(0)=-{758\over 45} \; ({\rm Eqs.} \; (1)-(4)) \; ,
\eqno (12)
$$
$$
\zeta(0)=-{241\over 90} \; ({\rm Eqs.} \; (5)-(7)) \; ,
\eqno (13)
$$
$$
\zeta(0)={89\over 90}-u-3u^{2}+{1\over 3}u^{3} \;
({\rm Eqs.} \; (8)-(11)) \; .
\eqno (14)
$$
By contrast, three-dimensional transverse-traceless (TT) 
perturbations yield $\zeta_{TT}(0)=-{278\over 45}$ in the cases
(12) and (13), and $\zeta_{TT}(0)={112\over 45}+3u-u^{2}
-{1\over 3}u^{3}$ instead of the result in Eq. (14).
It now remains to be seen what is the geometric form of 
heat-kernel coefficients when the Barvinsky boundary conditions
(5)--(7) are imposed in the de Donder gauge. The work in Ref. [2]
seems to show that this leads to infinitely many new universal
functions in Euclidean quantum gravity (such functions multiply
all possible local invariants in a linear combination whose
integration over $\partial M$ yields the boundary part of
heat-kernel coefficients).
 
If one wants to relate our analysis to the Lorentzian theory, one
has also to interpret the quantum state of the universe 
corresponding to the boundary conditions studied so far in
one-loop quantum cosmology. Moreover, the impossibility to
restrict the measure of the Euclidean path integral to
transverse-traceless perturbations raises further interpretative 
issues for the Lorentzian theory, where such a reduction to
physical degrees of freedom is instead quite natural. Last, but
not least, we find lack of one-loop finiteness of simple
supergravity on manifolds with boundary, if one of the two
boundary three-surfaces shrinks to a point [3]. Thus, many exciting
open problems remain also in the understanding of boundary
counterterms in supergravity theories.
\vskip 0.3cm
\noindent
\item {[1]} 
G. Esposito, A. Yu. Kamenshchik and G. Pollifrone,
{\it Euclidean Quantum Gravity on Manifolds with Boundary}
(Kluwer, Dordrecht, 1997).
\item {[2]}
I. G. Avramidi and G. Esposito, {\it New Invariants in the
One-Loop Divergences on Manifolds with Boundary} (old version
in HEP-TH 9701018).
\item {[3]}
G. Esposito and A. Yu. Kamenshchik, {\it Phys. Rev.}
{\bf D 54}, 3869 (1996).
 
\bye